\documentstyle[prl,aps,twocolumn]{revtex}
\newcommand{\be}{\begin{equation}}
\newcommand{\ee}{\end{equation}}

\begin{document}

\twocolumn[
\hsize\textwidth\columnwidth\hsize\csname @twocolumnfalse\endcsname

\title{ Effects of Geometric Phases in Josephson Junction Arrays }

\author{X.-M. Zhu$^1$, Yong Tan$^2$, and P. Ao$^1$ \\
$^1$Department of Theoretical Physics                 \\
Ume\aa{\ }University, S-901 87, Ume\aa, SWEDEN  \\
$^2$Department of Physics and Applied Physics \\
 University of Strathclyde, Glasgow G4 0NG, UNITED KINGDOM   }

\maketitle

\widetext

\begin{abstract}
We show that the {\it en route} 
vortex velocity dependent part of the Magnus force in
a Josephson junction array  is effectively zero, and predict
zero Hall effect in the classical limit.
However, geometric phases due to the finite superfluid density at 
superconductor grains have a profound influence on the quantum dynamics of
vortices.
Subsequently we find rich and complex Hall behaviors analogous to the
Thouless-Kohmoto-Nightingale-den Nijs effect in the quantum regime.
\end{abstract}

\noindent
PACS${\#}$s: 74.20.-z; 03.65.Bz; 72.15.Gd; 74.50.+r   

] 

\narrowtext


There have been extensive research activities on the 
vortex dynamics in Josephson junction arrays, 
where physical quantities which  determine the vortex dynamics,
such as the vortex potential, the effective vortex mass and viscosity,
are tunable by nanofabrication techniques.
One area which has started to attract attentions recently is the
Hall effect in Josephson junction arrays. 
In a homogeneous superconductor film it is known that 
the motion of a vortex resembles that of an electron in the presence of a
uniform magnetic field. 
The counter part of the Lorentz force for an electron
is the vortex velocity dependent
part of the Magnus force(hereafter called the transverse force)\cite{ao}.
Naturally, using the analogy for the fractional
quantum Hall effect in semiconductor heteorojunctions,
the existence of the quantum Hall effect in a 2-d Josephson junction array
has been argued by various authors\cite{stern,choi,odintsov}.
These proposals not only possibly have opened
a new practical way to utilize 
Josephson junction arrays, they also have a fundamental physical implication:
the realization of quantum Hall effect in Boson systems,
which can be used to test out our theoretical 
understandings.
In order to put the above attractive proposal on a firm
theoretical ground, a serious and thorough scrutiny should be conducted.
In the present paper we concentrate our attention on the role played by the
transverse force.
We have found that extreme cautions should be exercised when
using the transverse  force in Josephson junction arrays.
As a consequence, the results which we have obtained are different 
from those of early work\cite{stern,choi,odintsov}.

Our main results are the followings.
Because of a very large energy required, 
superconducting grains are inaccessible to vortices.
The vortex motion in a Josephson junction array is confined to the 
voids (nonsuperconducting areas) and tunneling barriers. 
Therefore the the local contact transverse
force is effectively zero in the classical limit.
However, in the quantum dynamics of 
the vortices, the transverse `force' does play an role
because of the Aharonov-Bohm type \cite{aharonov} scattering of 
the vortices by the superconducting grains. 
The phase of the wavefunction for a vortex will be influenced by a vector
potential linked to the finite superfluid density in the
superconductor grains. The vortices should  be considered as scattered by
a periodic array of the Aharonov-Bohm type fluxes and we have set up a
tight-binding hard-core  boson model to capture the main feature.
A straightforward way to find solutions for this 
boson model is to map it onto a fermion problem.
Rich quantum Hall behaviors are obtained following the
work of Thouless, Kohmoto, Nightingale, and den Nijs\cite{thouless},
In the following we present our analysis leading to above results.
Tunneling junctions and a square Josephson 
junction array will be assumed in the present 
paper. Our analysis can be carried over to other types of junctions and arrays
with necessary modifications.

We first show that the effect of the contact transverse force is zero, but
nevertheless geometric phases will be associated with the motion of vortices
in a Josephson junction array.
We will start from the nonlinear Schr\"odinger Lagrangian for the motion
of the superconducting condensate of a type II 
superconductor in the clean limit\cite{aitchison,stone}, 
because it contains all low energy, long wavelength dynamics 
at zero temperature, such as vortex dynamics and Josephson relations.
The nonlinear Schr\"odinger Lagrangian $L$ is
\be 
   L = i \hbar \psi^{\ast}\dot{\psi} - \frac{\hbar^2 }{2m^*}\nabla\psi^{*}
          \cdot\nabla\psi - V( \psi^{*},\psi ) \; ,
\ee
where $m^*$ is the Cooper pair mass, twice of the electron mass.
The Cooper pair wavefunction $\psi$ is normalized to the half of the superfluid
electron density. The charge of a Cooper pair is $2 e$. 
The coupling to the electromagnetic field can be put in. 
Since it will not influence our demonstration of the absence of the effect 
of the contact transverse force, it will not be written out explicitly.
The action is $S = \int dt d^2{\bf r} \; L(\psi^{*},\psi) $.
Variation of the action, $\delta S =0$, gives the
nonlinear Schr\"odinger equation for the condensate motion.
Writing the wavefunction $\psi$ as
\be
   \psi = \sqrt{\rho}\;\exp(i\theta) \; ,
\ee
 then nonlinear potential $V$ is given by
$   V = (\rho-\rho_{0})^{2}/2N(0) $,
where $N(0)$ is the density of states for each spin projection, and
$\rho_0$ the average Cooper pair density\cite{aitchison}.
Now we are ready to consider the motion of a single vortex 
in a homogeneous superconductor film.
The corresponding phase $\theta$ is a
function of the vortex position ${\bf r}_v(t)$, satisfying 
$\nabla\times \nabla\theta({\bf r}, {\bf r}_v) 
 =  2\pi q  \; \delta^2 ({\bf r}-{\bf r}_v) \hat{z} $.
Here $q=\pm 1$ is the sign of vorticity and $\hat{z}$ the unit vector
perpendicular to the film.  
We consider the vortex as a point particle,
and ${\bf r}_v(t)$ is defined as the
center of the vortex core. To obtain
a Lagrangian $L_v$ for the vortex, we will perform the
integration over ${\bf r}$ in the action 
$S = \int dtd^2{\bf r} \; L({\bf r},t,{\bf r}_v)$. 
Only the first term in the nonlinear Sch\"odinger Lagrangian, Eq.(1),
is relevant.  We begin by 
$
   \dot{\theta} ({\bf r}- {\bf r}_v)  = 
        \nabla_{{\bf r}_v} \theta ({\bf r}- {\bf r}_v)\cdot\dot{{\bf r}}_v 
$.
After the spatial integration, the first term in 
the Lagrangian $L$ in Eq.(1) gives rise to  a term in the vortex 
Lagrangian $L_v$ as 
$
   q{\bf A}_v\cdot \dot{{\bf r}}_v 
$
with 
\be
   q{\bf A}_v = - \hbar \int d^2{\bf r} \rho_0
       \nabla_{{\bf r}_v}\theta({\bf r}- {\bf r}_v) \; .
\ee 
Comparing with the known Lagrangian for  an electron  in
magnetic field 
$ 
   L_e = \frac{m}{2}\dot{\bf r}^2 +  e \; \dot{\bf r}\cdot{\bf A} 
$,
we conclude that the vortex is moving in a fictitious magnetic 
field. The `magnetic field'  for a vortex is identified as
\be
   {\bf B}_v =  \nabla _{{\bf r}_v}\times {\bf A}_v 
   =  - h \rho_0 \; \hat{z} \; .
\ee
Therefore the transverse force on a
moving vortex is given by
${\bf F}_m = q \; \dot{{\bf r}}_v \times{\bf B}_v $.

Next we consider a case in which the Cooper pair
density is smoothly modulated in space over the scale of the vortex core size, 
$\rho = \rho_0({\bf r})$. In such a case by repeating above derivation
we have
\be 
   {\bf B}_v({\bf r}_v)
   = - q \; \hbar \int d^2{\bf r} \rho({\bf r}) \nabla_{{\bf r}_v} 
      \times \nabla_{{\bf r}_v}\theta({\bf r}, {\bf r}_v)\; .
\ee
Generally, the phase $\theta({\bf r}, {\bf r}_v)$
in a inhomogeneous superconductor consists of two parts: rotational and
irrotational. The irrotational part will not contribute to the
above integral. The rotational part $\theta^r$ still satisfies 
\be 
   \nabla\times \nabla\theta^r({\bf r}, {\bf r}_v) =   2\pi q \; 
              \delta^2({\bf r}-{\bf r}_v) \; ,
\ee
which is required by the single valuedness of the wavefunction 
and the presence of a vortex.
Using the symmetry between ${\bf r}$ and ${\bf r}'$
in the above equation we obtain 
\be
   {\bf B}_v({\bf r}_v) 
   = - h \hat{z} \int d^2{\bf r}\rho({\bf r}) \delta^2 ({\bf r}-{\bf r}_v)
   =  - h  \rho_0 ({\bf r}_v) \hat{z} \; .
\ee
We notice that the fictitious magnetic field 
${\bf B}_v$ is propotional to the
Cooper pair density at  the vortex position. 
In other words the 
transverse  force on a vortex is a local property, just 
as the Lorentz force on an electron. 
Here the local Cooper pair density $\rho_0({\bf r})$ should be understood as
the average value over a regime much larger than the vortex core, but smaller
than the length scale for the density varying appreciably, the condition of 
the smooth variation.

Now we consider the extreme inhomogeneous case: a Josephson junction array.
If the superfluid density at some portions of the superconductor film is 
smoothly reduced to zero, voids are formed.  
According to Eq.(7) the local contact
transverse force is zero for a vortex at a void.
Voids are connected by low superfluid density regions, the tunneling 
barriers. 
The typical situation in a Josephson junction array is that 
the energy cost for a vortex to be at a tunneling barrier is much smaller than
the cost at a superconductor grain. 
Therefore vortices are confined to move on voids and tunneling barriers, 
an example of the guided vortex motion.
Although the transverse force is finite on a tunneling barrier,
its effect on the classical vortex motion is zero,
because vortices must move along tunneling junctions.
To summarize above analysis, the
local contact transverse force does not play a role in vortex dynamics
in a Josephson junction array. Subsequently in the classical limit 
the transverse force can be ignored.
We note that our above conclusion differs with the one 
in a recent preprint\cite{gaitan}, where no such energetic constraint has been
applied\cite{horovitz}.
The absence of the {\it en route}
transverse force is in agreement with  experimental observations 
on the classical vortex dynamics 
in Josephson junction arrays at relatively
high temperatures and with  large effective masses such as
vortices moving perpendicular to the driving current\cite{vertical}, 
no detectable Hall effect\cite{nohall}, 
and the straightline balistic motion\cite{balistic}.
The condition for the classical limit will be given at the end of the paper.

We turn to the Hall effect in the quantum regime.
Zero transverse force itself does not necessarily 
exclude all possible quantum Hall effects.
The difference between the
classical and quantum dynamics lies in the role played by the phase,
such as the geometric phase associated with the vector potential 
described in Eq.(3),
where the vector potential can be finite although the `magnetic field' is zero.
This is precisely the case of the Aharonov-Bohm effect which illustrates that 
in quantum mechanics potentials are more fundamental than 
forces\cite{aharonov}.
The analogy of the Aharonov-Bohm effect has been studied in Ref.\cite{zhu}. 
In the remaining part of the paper we explore the logical consequences
of the geometric phases for the quantum Hall effect
in a Josephson junction array. 
In order to reveal the essential physics and to gain the physical insight, 
we will idealize our problem and approximate vortices as hard-core bosons.

To be commensurated with the existence of the vortex inaccessible
regions and the geometric phases,
we consider the the tight-binding limit of vortex motion. 
The corresponding Hamiltonian may be written as
\be
   H= t \; \sum_{(l,j)} a^{\dag}_l a_j \; e^{i A_{lj}}
        + \sum_{l,j} a^{\dag}_l a_l \; V_{lj} \; a^{\dag}_j a_j \; ,
\ee
where $ a_l $ is the boson annihilation 
operator for a vortex at j-th void, and 
$( \; )$ stands for the summation over nearest neighbors.
The phase $A_{lj}$ is
defined on the links connected the nearest neighbors, and its 
sum around a plaquette is equal to the geometric phase $2 \pi \phi_0$:
$\sum_{plaquette} A_{lj} = 2 \pi \phi_0$. A uniform
geometric phase  will be assumed, where the number of `fluxes'  $\phi_0$ 
is the number of Cooper pairs on a superconductor grain, which 
may be controlled by a gate voltage.
The interaction between vortices is described by $V_{lj}$, which is long range 
and repulsive. We will treat it as a short range repulsive interaction for
a first approximation. 
The tunneling matrix element $t$ is, in terms of the parameters for 
a Josephson junction, an order of $\sqrt{E_J E_C}
\exp\{ - O(1) \sqrt{E_J/E_C} \}$, 
where $E_J$ 
is the Josephson junction energy and $E_C$ the junction charging 
energy\cite{theor}. 
The energy scale for the repulsive interaction is $E_J$\cite{theor}, 
which is much larger than the tunneling matrix element $t$.
Nevertheless with a considerable amount of energy two vortices 
can be put on one position. This suggests that vortices are hard-core bosons. 
Then we may approximate the vortex problem described by Eq.(8) as 
a hard-core boson problem, an approximation has already been implemented 
in Refs.\cite{choi,odintsov}.

We are ready to discuss the quantum Hall effect of the idealized vortex
problem. We do this by mapping the hard-core boson problem onto a 
fermion problem by attaching odd number of `fluxes' on each vortex.
This is a standard procedure.\cite{fradkin}
The resulting Hamiltonian for the fermion problem is 
\be
   H= t \; \sum_{(l,j)} c^{\dag}_l c_j \; e^{i[A_{lj} + {\cal A}_{lj} ]} \; , 
\ee
where $c_j$ is the corresponding the fermion annihilation operator at the j-th
void.
The number of statistical fluxes $\phi_s$ 
at the j-th void satisfys the constrain
$  \phi_s = - (2m + 1)  < c^{\dag}_j c_j > $, with 
$\sum_{plaquette} {\cal A}_{lj} = 2\pi \phi_s$, 
which means that $2m + 1$ fluxes have been attached to each vortex.
We assume that this mapping will give a mean field solution with an energy
gap separated from its excitations. 
Then the statistical fluxes can be adiabatically smeared over the lattice and
effectively detached from vortices, as shown in Ref.\cite{laughlin}.
In this case $\phi_s = - (2m + 1) n $, with $n$ is the magnetic flux 
frustration, the number of vortices per plaquette. 
Then the  resulting mean field problem is exactly 
the Harper-Azbel-Wannier-Hofstadter problem, where energy gaps do exist.
The quantum Hall behaviors of such a problem have been studied in detail
by Thouless, Kohmoto, Nightingale, and den Nijs\cite{thouless}.
For such a system the quantum Hall conductance $\sigma^f_H$ is
\be
   \sigma^f_H = t_r \; ,
\ee
with the integer $t_r$ the solution of the Diophantine equation
\be
   r = s_r q + t_r p \; .
\ee
Here the number of fluxes per plaquette $\phi = \phi_0 - \phi_s = p/q$, 
with $p$ and $q$ coprime,
$n = r/q$, and $r$, $s_r$, $t_r$ integers with $|t_r| \leq q/2$.
Remember that the mapping has generated a Chern-Simons term, which has
a contribution to the Hall conductance as 
\be
   \sigma^s_H = \frac{1}{ 2m + 1 } \; ,
\ee
the quantum Hall conductance of the original vortex system is
then\cite{fradkin,read}
\be
   \frac{1}{\sigma^v_H } = \frac{1}{\sigma^f_H } + \frac{1}{\sigma^s_H } \; .
\ee
Converting back into the electric Hall conductance and putting back the unit,
we find the electric quantum conductance of the Josephson junction array is
\be
   \sigma_H = \frac{4e^2} {h} \; \frac{1}{\sigma^v_H } \;.
\ee
It is interesting to observe that a boson-to-fermion mapping has been used here
to find the incompressible quantum Hall states, while in the usual
fractional quantum Hall effect it is the other way around\cite{zhang}.
As well known in the previous study of quantum Hall effect\cite{read,fradkin}
for a given set of the `flux' $\phi_0$ and the frustration $n$, 
there may exist several values of $m$, that is, several mappings,
with their mean-field solutions all corresponding to filled bands which are
separated from excitations by energy gaps.
If such a case occurs, detailed calculation is needed fo find the $m$ with
the largest energy gap, which is the most stable one.

One can check that following symmetries hold for the quantum Hall conductance
$\sigma_H$:  the periodicity,
$\sigma_H(\phi_0, n) =   \sigma_H(\phi_0 + 1, n)$ ;
the odd symmetry, $\sigma_H(\phi_0, n) = - \sigma_H(-\phi_0, n)$; 
the particle-hole symmetry, $\sigma_H(\phi_0, n) = - \sigma_H(\phi_0, 1 - n)$.
Because $\sigma^f_H$ is a non-monotonic and rapidly varying function of
the `flux' (number of Cooper pairs per plaquette) 
$\phi_0$ and the frustration $n$, so will be $\sigma_H$.
Particularly, both positive and negative Hall conductance may be easily 
reached.
For example, for $n = 1/5$ and $\phi_0 = 1/3$, we find that
$\sigma_H = \frac{10}{3}\frac{4e^2} {h} $ with $m=1$; and 
for $n = 1/3$ and $\phi_0 = 1/5$, 
$\sigma_H = - \frac{6}{5} \frac{4e^2} {h} $ with $m=-1$.
This is in sharp contrast with the previous proposal of the quantum 
Hall effect in a Josephson junction array\cite{stern,choi,odintsov}.
There are special sets of $\phi_0$ and $n$ such that 
$2m \; n = \phi_0 $, that is, 
in a boson-to-boson mapping the statistical flux cancels the real flux.
In this case the mean-field solution is automatically within a gap, and the
Hall conductance is
\be
   \sigma_H = 2m \; \frac{4 e^2 }{h}  \; .
\ee
This is to the contrast with the fermion-to-fermion mapping case discussed in 
Ref.\cite{halperin}, where there is no energy gap which 
separate the mean-field
solution and excitations.
With these specific sets of $\phi_0$ and $n$ and in the zero limit 
of their fraction parts 
one can take the continuous limit of the tight-binding 
model\cite{fradkin} to
recover previous proposed quantum Hall effect based the contact
transverse force\cite{stern,choi,odintsov}.

Two comments concerning previous work are also in order.
First, the  conditions to observe
the Aharonov-Casher effect in a Josephson junction array
have been discussed in Ref.\cite{zhu}. Because of the extensive
spreading of the associated magnetic field,
they are unlikely to be fulfilled, which applies to the
situation considered in Ref. \cite{choi}.
Second, in Ref.\cite{odintsov}
the quantum Hall effect was argued from the dynamics of Cooper pairs (charged
bosons), a dual picture of the vortices.  
Our comment here is that even in this dual picture 
the contact Lorentz force is not responsible for the
quantum Hall effect, 
but the real Aharonov-Bohm phase similar to the vortex 
picture as discussed in the proceeding paragraphs is.
 
To summarize, the effect of the vortex velocity dependent part of the
Magnus force is found to be zero in the vortex dynamics 
in a Josephson junction array. Instead the geometric phases are 
important in the vortex quantum dynamics.
We predict that in the classical limit there is no Hall effect at all,
but in the
quantum limit rich quantum Hall behaviors should exist:
positive and negative Hall conductances
determined by the Thouless-Kohmoto-Nightingale-den Nijs effect and the 
Chern-Simons contribution.
The relevant energy scale is the tunneling matrix element $t$.
For a Josephson junction energy $E_J \sim 1$ K, and the junction charging 
energy $E_C$ with a value such that $E_J /E_C \sim 1$, we find 
$t \sim 100$ mK.
When the temperature is higher than $t$, thermal fluctuation will destroy the
quantum coherence and the vortices move classically.
The quantum regime is realized for temperatures lower than $t$ where 
the phase coherence is preserved. Experimentally the
quantum regime should be accessible in principle and a
recent report has already 
shown the existence of the Hall effect 
at low temperatures\cite{chen}.

{\bf Acknowledgements}:
We thank Petter Minnhagen and Jens Houlrik for useful discussions, 
and Frank Gaitan for sending us his preprint of Ref.\cite{gaitan} and 
following up discussions.
We also thank Per Delsing for showing the raw data in Ref.\cite{chen} 
which motivated present study, and Chiidong Chen and 
David Haviland for discussions, as well as Gerd Sch\"{o}n for pointing out
a misprint in the tunneling matrix element in the manuscript.
This work 
was supported in part by Swedish Natural Science Research Council.

\end{document}